\documentclass[]{spie} 

\usepackage[]{graphicx}

\title{Heterogeneities and local fluctuations in glassy systems} 

\author{Leticia F. Cugliandolo
\skiplinehalf
Laboratoire de Physique Th\'eorique, Ecole Normale Sup\'erieure, \\
24 rue Lhomond, 75231 Paris Cedex 05, France, \\
Laboratoire de Physique Th\'eorique et Hautes Energies, Jussieu, \\
4 Place Jussieu 75252 Paris Cedex 05, France
}

\authorinfo{E-mail: leticia@lpt.ens.fr}

\begin{document} 
  \maketitle

\begin{abstract}
The aim of this report is to review a theoretical approach that has
been proposed recently to describe dynamic fluctuations in glassy
systems (work in collaboration with H. E. Castillo, C. Chamon,
P. Charbonneau, J. L. Iguain, M. P. Kennett, D. R. Reichman and
M. Sellitto).  Firstly, I summarize some of the main features of the
averaged and global non-equilibrium relaxation of glassy systems,
weakly sheared viscous liquids and weakly tapped granular matter, and
how these results have been successfully reproduced with a
mean-field-like analytic approach. Secondly, I explain the outcome of
more refined experimental and numerical measurements that point at
examining the dynamics at a mesoscopic scale, and the role of noise
measurements in this respect. Finally, I discuss how our theoretical
approach can be confronted to new experimental and numerical tests.
\end{abstract}



\section{Introduction}
\label{sect:intro}  
A system may not be able to reach equilibrium with its environment for
several reasons.~\cite{Cu02} The most common cases are those in which
the time needed to equilibrate the sample falls beyond the
experimental time-window.  Examples are given by domain growth, phase
separation and classical and quantum glassy systems.  On the other
hand, external forces can drive a sample out of equilibrium, as in
rheological measurements. If the systems tested have a tendency to
evolve slowly, and if the perturbations used are not too strong, their
slow dynamic behaviour may resemble strongly the one of glasses. This
is the case, for instance, of sheared super-cooled liquids, and weakly
tapped granular matter.

Even if there is a kind of universality in the form of the decay of
bulk observables in glassy systems, its origin might vary
significantly from material to material.  In some systems it is pretty
easy to identify the mechanism behind the dynamic arrest and the slow
non-equilibrium dynamics at low temperatures. For instance, a
ferromagnetic spin model with quenched random fields has a very slow
coarsening dynamics since the interfaces get pinned by disorder. In
more generic glassy materials, the impressive slowing down
demonstrated by the growth of the relaxation time by several orders of
magnitude when approaching the glass transition is not accompanied by
any detectable change in structure.  Indeed, establishing a relation
between dynamics and structure in glassy materials has remained
elusive.

The recent development of powerful experimental
techniques~\cite{Ediger,Kegel,Weitz,Weeks1,Weeks2,Nathan,Miller,Sergio2,Luca2}
and the use of extensive numerical
simulations~\cite{Laird,2d,Donati99,Heuer,Johnson,Yamamoto,Starr02,Ogli,Vollmayr,Caetal02,Caetal02-2,Chetal04}
are starting to yield very interesting information about the
dynamic behavior at a mesoscopic scale. In these studies, emphasis is set on
identifying the most relevant rearrangements that take place during
evolution.  In addition, a detailed analysis of temporal fluctuations
in the evolution of the relevant observables yields complementary
information that allows one to distinguish between intermittent and
continuous dynamics, the former being associated with sudden, large
rearrangements.~\cite{Weeks2,Sergio2,Luca2} The insight obtained from
these studies might be very valuable to understand the reason for the
glassy arrest and to develop a complete theoretical description of it.
The examples cited in the first paragraph belong to a particular class
of out of equilibrium systems: those with a {\it very slow
dynamics}.~\cite{Cu02} Exploiting this property, as well as other more
subtle ones,~\cite{Kithwo,Cuku93,Cuku94,Bocukume,Cuku-japan,Chetal02}
one can hope to develop a common theoretical description of all of
them.  Indeed, the decay of different averaged two-time global
correlation functions in rather diverse glasses have a very similar
pattern. For example, the averaged two-time correlations between
global magnetic fluctuations in disordered spin systems~\cite{Didier},
and the averaged two-time intensity-intensity correlations in light
scattering experiments in colloidal suspensions decay with a very
similar form~\cite{Luca2,Bonn,Virgile,Luca1}, see
Fig.~\ref{fig:corr}. In the paramagnetic or super-cooled liquid
(stationary) and the spin-glass or jammed (aging) phases these are
rather well described analytically by the solution to the dynamics of
fully-connected disordered spin models~\cite{Kithwo,Cuku93}. In the
late 80s Kirkpatrick, Thirumalai and Wolynes~\cite{Kithwo} showed
that, above the glass transition, these models lead to dynamic
equations for the disorder and statistical averaged two-time functions
that are identical to those derived within the mode-coupling approach
to super-cooled liquids~\cite{Gotze}.  More recently it has been
realized that these models extend the mode-coupling approach below the
glass transition~\cite{Frhe,Bocukume,Latz}.  The precursor two-step
relaxation of averaged bulk correlations in super-cooled liquids, the
growth of the structural relaxation time when lowering the external
temperature, aging effects in the glassy phase, shear thinning
effects, etc. are captured by these models~\cite{Cu02}.

One of the more interesting and open theoretical problems now is how
to include in such a description large-scale fluctuations (in space
and time) that are possibly associated to cooperative and correlated
rearrangements of particles.

In equilibrium and away from criticality any global observable of a
macroscopic system has Gaussian fluctuations. At criticality, instead,
one observes non-Gaussian fluctuations due to the divergence of the
correlation length and the non applicability of the central limit
theorem. Still, scale invariance at the critical point constrains the
possible probability distributions; these are determined by the
universality class to which the systems belong.  A similar criterium
to classify the probability distributions of the fluctuations of
macroscopic observables in critical nonequilibrium systems that is
based on the use of symmetries, has been proposed~\cite{Zoltan}. The
glass transition, where the system falls out of equilibrium, is a
dynamic crossover and neither a dynamic nor a thermodynamic
transition. Even if, strictly speaking, there is no phase transition,
the dynamics in the {\it full} glassy phase shares some features with
usual critical dynamics. Let us mention three properties that give
support to this statement. The correlation functions, say for fixed
waiting-time as a function of the total time, do not relax
exponentially but decay in a much slower manner. In the free-energy
landscape the dynamics at long-times can be associated to a wandering
along flat directions (times are taken to be long but not so long as
to allow for jumps over barriers in the free-energy landscape that
scale with the size of the system). This property has been made
precise for the solvable models mentioned above~\cite{Cuku93,Kula,Bi}
and it has been shown under certain assumptions for finite dimensional
disordered models~\cite{Chetal02}. A dynamic correlation length is
expected to reach a power law dependence at sufficiently long times
belonging to the same regime~\cite{Caetal02-2} (see also~\cite{corr-length}).  
Thus, adapting arguments usually used in the study of critical phenomena, one can
expect that the form of the probability distribution functions of
fluctuating dynamical quantities will be constrained by the symmetries
in the problem. On the experimental and numerical side, these
distributions can be measured and classified. Clearly, noise
measurements play a very important role in this respect. On the
theoretical side, one can construct effective sigma models for the
fluctuating quantities, derive from them scaling laws for the
fluctuations~\cite{Chetal02,Caetal02,Caetal02-2} and, in certain
cases, the actual form of the scaling functions~\cite{Chetal04}.  This is
the route we have followed recently.

In this presentation I review several aspects of the non-equilibrium
dynamics of glassy systems, including aging relaxational phenomena and
rejuvenation caused by external perturbations.
Until recently, experiments, numerical studies and theory have focused
on the behavior of averaged bulk quantities.  The interest is now
shifting to the study of dynamic heterogeneities, and
intermittency. The aim is to detect similarities and differences in
the behavior of temporal and spatial fluctuations in different glassy
systems.  I present a short summary of the clearest experimental
results concerning fluctuations and I summarize the theoretical
approach that we developed to describe
them.~\cite{Caetal02,Caetal02-2,Chetal04,Chetal02}

\section{Nonequilibrium relaxation in glassy systems}

Out of equilibrium relaxational dynamics occurs, for instance, when
one suddenly quenches a system with ferromagnetic interactions below
its Curie temperature. The system evolves from the very disordered
initial condition via the growth of domains of up and down magnetic
order. Since the typical domain radius behaves as $t^{1/2}$ the time
needed to order a sample of linear size $L$ goes as $L^2$ and cannot
be reached with finite times with respect to $L$. When quenched
disorder, as random fields, are also present the dynamics keeps the
same characteristics but it is much slower~\cite{Levelut}.

In the domain growth example one easily visualizes the growth of
order. Going across the super-cooled liquid--glass crossover thorough
any path (annealing in temperature, crunch in pressure, etc.)  the
systems fall out of equilibrium as demonstrated by several facts: the
observables depend on the preparation used, the systems age, etc.
Whether there is a growing order controlling their evolution in the
glassy phase is still an open question. Glasses of very different
types have been identified and studied and we mention only some
samples that we shall discuss in this report: dipolar
glasses~\cite{Levelut},
spin-glasses~\cite{noise-SG-Paris,noise-SG,Didier}, colloidal
suspensions~\cite{Bonn,Virgile,Luca1,Luca2,Kegel,Weitz,Weeks1,Weeks2,Sergio1,Sergio2},
polymer glasses~\cite{Sergio1,Sergio2,Nathan},
glycerol~\cite{Miller,Tomas} and relaxor ferroelectrics~\cite{Weiss}.

\begin{figure}
\begin{center}
\begin{tabular}{c}
\includegraphics[height=4.3cm]{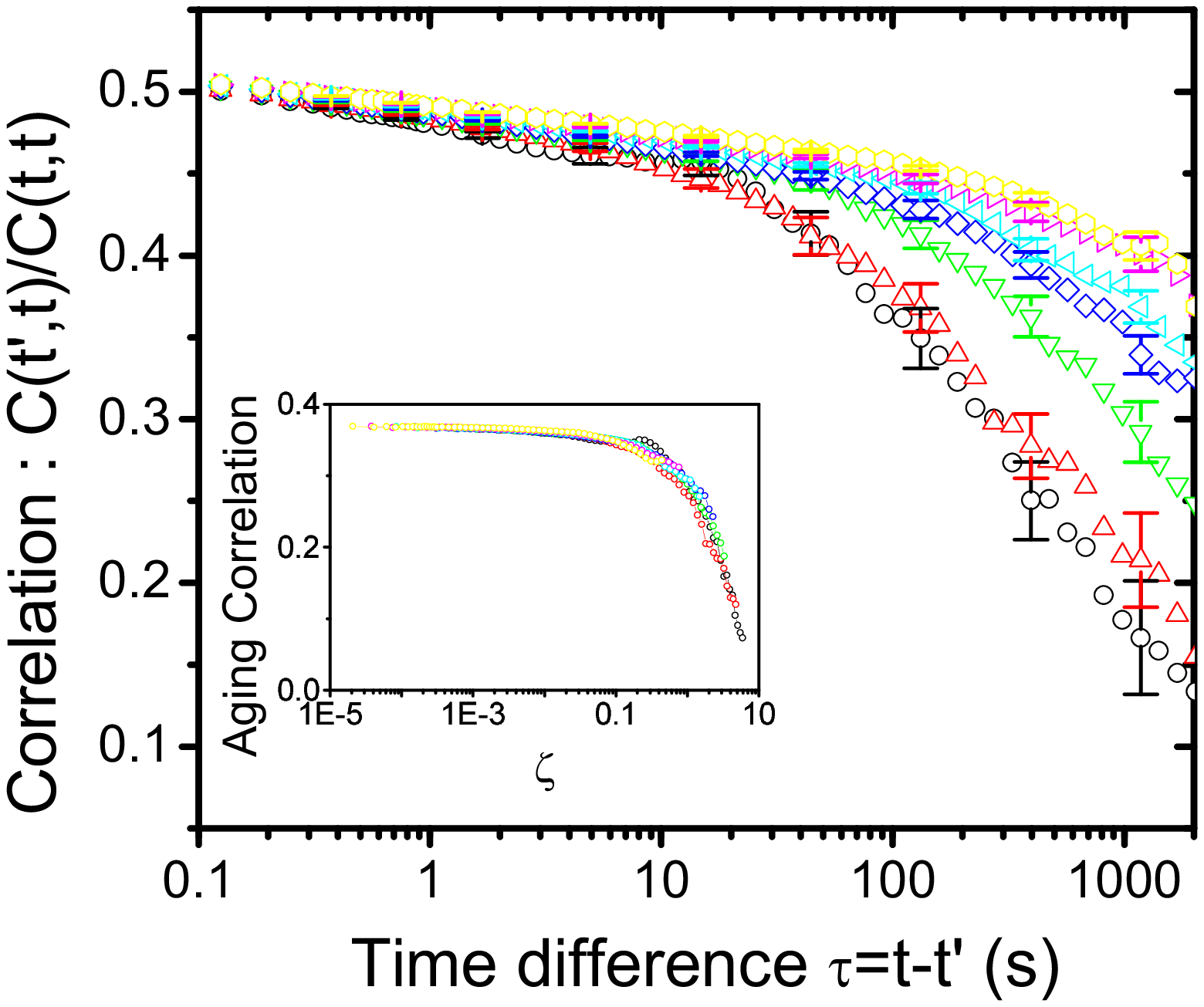}
\hspace{1cm}
\includegraphics[height=4.3cm]{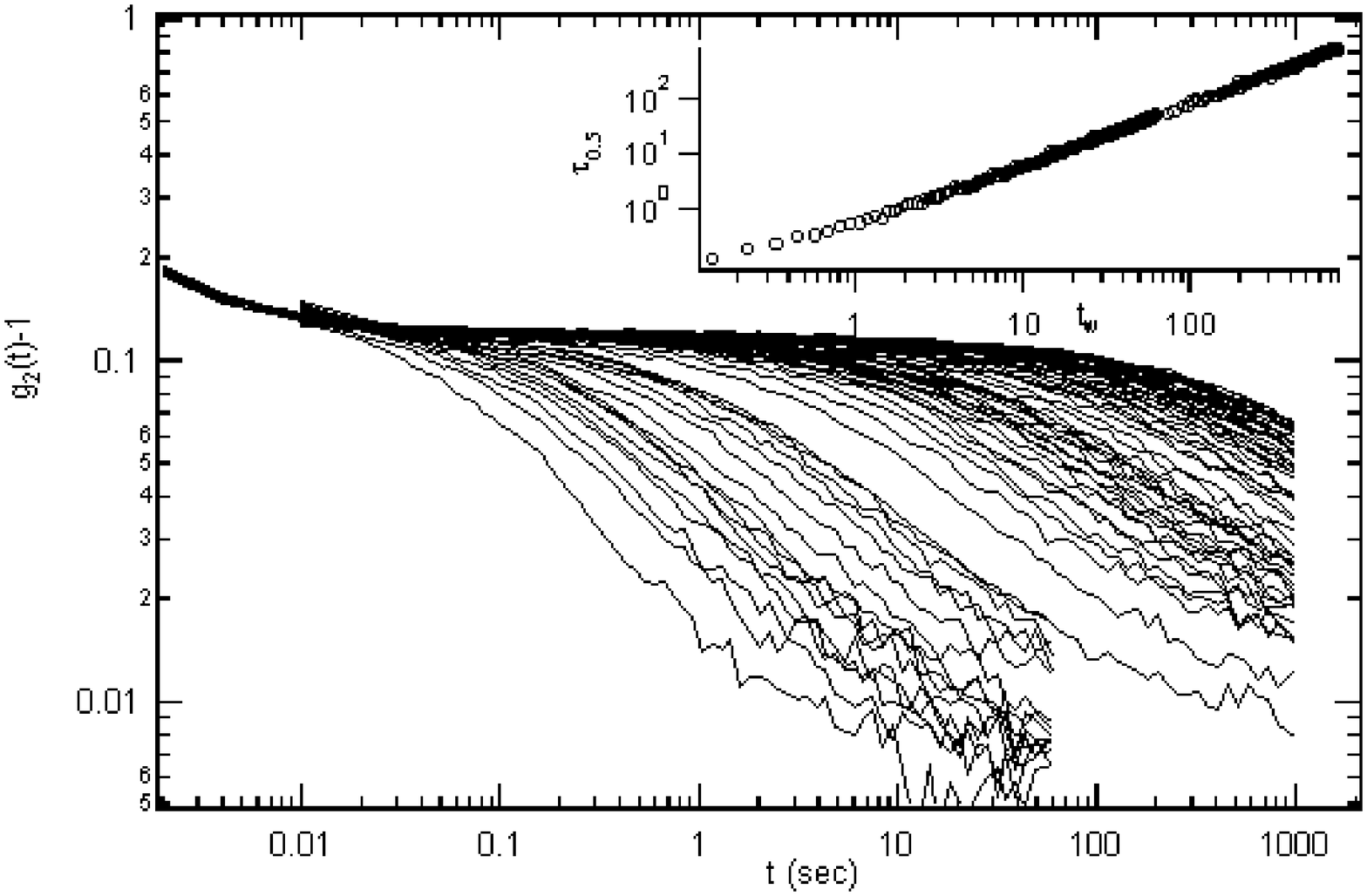}
\end{tabular}
\end{center}
\caption[example] 
{ \label{fig:corr} Left: Decay of the normalized two-time correlations
between magnetic fluctuations~\cite{Didier}, $m_\alpha(t)$, in the
insulating spin-glass CdCr$_{1.7}$In$_{0.3}$S$_4$ (the index $\alpha$
indicates the number of the experimental run). The magnetic
fluctuations are measured using a short temporal coarse-graining. The
correlation $C(t,t_w)$ is obtained by averaging over ${\cal N}$
repetitions of the same experience the product of $m_\alpha$ at two
subsequent times: $C(t,t_w) \propto {\cal N}^{-1}
\sum_{\alpha=1}^{\cal N} m_\alpha(t) m_\alpha(t_w)$.  Each curve
corresponds to a fixed waiting-time and it represents $C$ as a
function of the time-difference.  The lowest (uppest) curve
corresponds to the shortest (longest) waiting-time. Right: Decay of
the normalized intensity-intensity correlation function,
$g_2(t,t_w)-1$, measured with light scattering on a colloidal system
in its glassy phase~\cite{Virgile}. The curves are averaged over
different runs and using a temporal coarse-graining. The same system
under a permanent shear presents very similar decorrelation with the
unique difference that the second decay below the plateau is now
stationary with a characteristic time determined by the shear
strength.\cite{Cukulepe,Yamamoto,Berthier}}
\end{figure}

The above examples concern systems that are not able to reach
equilibrium with their environments in a reasonable time but that, let
evolve on astronomical time-scales, will eventually equilibrate.  A
liquid can be driven to a slow out of equilibrium stationary regime by
a weak shear. This is a force that does not derive from a potential
and its effect cannot be described with a usual statistical mechanics
approach. If the liquid is sufficiently dense, its sheared dynamics
resembles the one of a purely relaxing super-cooled liquid (see
Fig.~\ref{fig:corr} for the details) but it occurs out of
equilibrium~\cite{Bonn,Virgile,Yamamoto,Berthier}.

Another family of materials that have captured the attention of
experimentalists and theoreticians in recent years is granular
matter~\cite{Nagel,aging-granular,dAnna}.  Since the potential energy
needed to displace a macroscopic grain by a distance equal to its
diameter, $m g d$, is much larger than the characteristic thermal
energy, $k_BT$, thermal activation is totally irrelevant for systems
made of macroscopic grains. Therefore, in the absence of external
driving granular matter is blocked in metastable states. There is no
statistical mechanics approach capable of describing its static
behavior.  Instead, when energy is pumped in the form of shearing,
vibration or tapping, transitions between the otherwise metastable
states occur and granular matter slowly relaxes towards configurations
with higher densities.
Glassy features such as hysteresis as a function of the amount of
energy injected, slow dynamics, and non stationary correlations have
been exhibited.

Aging means that older systems relax in a slower manner than younger
ones. One defines the age of a system as the time spent in the phase
under study. For instance, the age of a system that is suddenly
quenched from high-$T$ to low-$T$ is simply the time elapsed since the
quench. Colloidal suspensions at a given concentration are usually
initialized by applying a strong stirring that is suddenly stopped at
the initial time.  The aging properties are studied by monitoring the
time evolution of correlation and response functions. In the former
experiments one lets the system evolve and compares its configuration
at a waiting-time $t_w$ with the one reached at a subsequent time
$\tau+t_w$ (see Fig.~\ref{fig:corr}).  In the latter one perturbs the
system at $t_w$ with, {\it e.g.} a dc or an ac field, and follows the
evolution of the linear response to the perturbation.  These are {\it
global or bulk} functions in the sense that they represent the
dynamics of the full system, and they are averaged over different
repetitions of the experiment (statistical average) and sometimes a
coarse-graining in time using a short time-window around the
observation instant is also implemented. In the glassy phase both
correlations and responses depend on $t_w$ in an aging manner and,
within the experimentally accessible time-window, this trend does not
show any tendency to stop. At temperatures that are close but above
$T_g$, or concentrations that are close but below $\phi_g$, one
observes ``interrupted aging'' that is, a dependence on the age of the
system until it reaches the equilibration time ($t_w > t_{\sc eq}$)
where the dynamics crosses over to an equilibrium one. In equilibrium
correlation and response measurements are related in a system
independent manner by the fluctuation-dissipation theorem.  Out of
equilibrium this general relation does not hold and important
information can be extracted from its modifications~\cite{Cu02} (see
Sect.~\ref{Teff}).

Aging has an easy interpretation within coarsening
systems.~\cite{Cu02} The motion of interfaces is usually driven by
their curvature.  Since domains grow, the curvature of the domain
walls decreases in time and the dynamics slows down as time elapses.
Comparing the configuration at $t_w$ and at a later time $\tau+t_w$,
one finds a clear separation of time-scales depending on the relative
value of $\tau$ with respect to $t_w$. For short time differences with
respect to a characteristic time $\tau_0(t_w) \equiv 1/[d_{t_w} \ln
{\cal R}(t_w)]$, with ${\cal R}(t_w)$ the typical domain radius at
$t_w$, domain walls do not move. Then, if the ratio between surface
and volume of the domains vanishes in the thermodynamic limit (a
hypothesis that might be violated in some systems), the overlap
between the configurations at $t_w$ and $\tau+t_w$ simply takes into
account the thermal fluctuations within the domains and the
correlation decays as in equilibrium from $1$ at equal times to a
value called $q_{\sc ea}(T)$ [that equals $m^2_{\sc eq}(T)$ in a
ferromagnetic system] when $\tau\to\tau_0(t_w)$. For time-differences
beyond $\tau_0(t_w)$, the correlations decay below $m^2_{\sc eq}(T)$
since the interfaces move and one compares configurations with very
different domain structures.  As already mentioned, coarsening occurs,
for instance, in ferromagnets below $T_c$. The aging properties of the
orientational glass K$_{1-x}$Li$_x$Ta0$_3$ have also been interpreted
using such a picture~\cite{Levelut}.

In structural glasses, a pictorial explanation of aging is also
possible imagining that each particle sees a cage made of its
neighbors.~\cite{Cu02} When $\tau$ is short compared to a
characteristic time $\tau_0(t_w)$ each particle typically rattles
within its cage and the decorrelation is only characterized by thermal
fluctuations.  As in the coarsening example, the correlations decay in
a stationary manner from its value at equal times to $q_{\sc ea}(T)$
in this time regime.  When $\tau$ increases, the motion of the
particles destroys the original cages and one sees the structural
relaxation.  The waiting-time dependence implies that the cages are
stiffer when time evolves. The motion of a tagged particle observed
with confocal microscopy demonstrated this scenario~\cite{Weeks1}.

By shearing a liquid one usually facilitates its flow and the
relaxation time decreases with increasing shearing rate (shear
thinning).  By shearing a glass one typically introduces a
characteristic time that corresponds to the longest relaxation
time. Thus, aging is interrupted for longer waiting-times and the
sample is rejuvenated by the external
perturbation~\cite{Bonn,Virgile,Cukulepe,Yamamoto,Berthier}. The
two-step structure of the relaxation remains unaltered when a weak
shear is applied but the scaling in time is, however, modified. This
observation is key to the derivation of the theory of fluctuations we
shall discuss in Sect.~\ref{theory}.

The effect of an oscillatory drive, typically used to provoke the 
relaxation of granular matter, can be rather different. Indeed, 
the study of mean-field models of granular matter indicated that 
aging might not be arrested by such a perturbation.~\cite{Becuig}  This is 
the subject of current experimental and numerical 
investigations~\cite{aging-granular}.

The curves in Fig.~\ref{fig:corr} can be scaled in a quite satisfactory 
way by using the following form:~\cite{Cu02,Cuku93,Cuku94}
\begin{equation}
C(t,t_w) = C_{\sc fast}(t-t_w) + C_{\sc slow}(t,t_w)=
f_{\sc fast}\left(\frac{h_{\sc fast}(t)}{h_{\sc fast}(t_w)} \right) 
+ f_{\sc slow} \left(\frac{h_{\sc slow}(t)}{h_{\sc slow}(t_w)}\right)
\label{global-C}
\end{equation}
with $h_{\sc fast}(t) = e^{t/\tau}$ and $h_{\sc slow}(t)$ a
system-dependent monotonic function. In aging systems the first term
describes the stationary approach to the plateau, the second term the
$t_w$-dependent departure from it.  In some cases, as the $3d$ {\sc
ea} model, one finds that $h_{\sc slow}(t)=t$ describes rather
accurately the available data. In some other aging systems like the
insulating spin-glass studied by H\'erisson and Ocio~\cite{Didier} an
enhanced power law does a better job.  Notably, the effect of a weak
shear is to modify $h_{\sc slow}(t)$ rendering it an exponential in
such a way that the second decay becomes stationary as well. It only
weakly modifies instead other important features of the relaxation as
the value of the plateau or how the {\sc fdt} (see Sect.~\ref{Teff})
is modified.  In theoretical terms, the important response of the
scaling function $h_{\sc slow}$ to external perturbations is due to
the time-reparametrization invariance of the dynamic action for
the structural relaxation.~\cite{Cuku-japan,Chetal02} The consequence
of this symmetry as fluctuations are concerned will be explained in
Sect.~\ref{theory}.

\section{Heterogeneous dynamics: experiments and simulations}
\label{exps}

Two extreme interpretations of the origin of the non-exponential
relaxation in super-cooled liquids are the so-called ``heterogeneous''
and ``homogeneous'' scenarios.~\cite{Ediger} In the former, the
non-exponential relaxation is supposed to be due to the exponential
decay of independent cooperative regions (clusters of estimated 2-4 nm
typical size) each with its own characteristic time. Note that any
well-behaved relaxation function can be expressed as
$R(t)=\int_0^\infty d\tau \, g(\tau) \, e^{-t/\tau}$. The very
non-trivial ingredient in this scenario is that the exponentials are
associated to the relaxation of individual and localized regions in
the sample.  In the latter scenario instead the sample behaves
homogeneously and the decay is intrinsically non-exponential. Clearly,
any intermediate scenario is also  possible, with some regions
transiently decaying exponentially, and later returning to the bulk
behavior.

In the descriptions discussed in the previous paragraph the
heterogeneities are supposed to have an exponential relaxation. More
generally, an heterogeneity can be a region in the sample that relaxes
very differently from the bulk but not necessarily with an exponential
function.  Clearly, the heterogeneities will not be static and they
will appear and disappear as dynamic fluctuations. Thus, we shall
adopt a more generic definition of dynamic heterogeneity that is ``any
mesoscopic region in the sample that transiently relaxes with a very
different law from the one in the bulk''.

The existence of dynamic heterogeneities in super-cooled liquids and
glasses has been suggested on the basis of experiments performed using
many different techniques.~\cite{Ediger} Some of these methods test
the heterogeneous character in a very indirect manner.~\cite{Joseluis}
Others, have an easier interpretation and yield a more direct probe of
the structure and dynamics.  In this Section we summarize the evidence
for heterogeneous dynamics in super-cooled liquids and glasses
focusing only on the results obtained with light-scattering,
atomic-force probes, and confocal microscopy.  We also briefly review
the numerical simulations done with the purpose of studying dynamic
heterogeneities.  Most of these works have concentrated on the
super-cooled liquid phase of a binary Lennard-Jones
mixture~\cite{Laird,2d,Donati99,Heuer,Johnson,Yamamoto} and polymer
melts~\cite{Starr02} in two and three dimensions and the
heterogeneities have been studied by tagging each particle and
classifying them according to their mobility during a selected
time-interval.  Some recent papers treat the glassy phase of the
particle system in a similar way.~\cite{Vollmayr}
 

Following the dynamics of individual particles is very useful to get
an intuitive understanding of the evolution of the system but it is
hard to use as a direct input in a theoretical approach. These
results, as well as the numerical analysis of spin
models~\cite{Caetal02,Caetal02-2}, suggest to use a coarse-grained
description of the local properties. We shall discuss this
approach~\cite{Caetal02,Caetal02-2,Chetal02} in Sect.~\ref{theory}.

\subsection{Spatial heterogeneities}

Israeloff and collaborators used non-contact scanning probe microscopy
techniques (cantilevers)~\cite{Nathan} to measure the fluctuations in
the dielectric properties of various polymer films at a mesoscopic --
nanometric -- scale. Working at and below their glass transition these
authors found a polarization noise power spectrum that is typically
$1/f^\gamma$ but whose form or exponent vary in time.  This variation
was associated to transient dynamical heterogeneities with lifetime
similar to the $\alpha$-relaxation time. Interestingly enough, the
appearance of smooth Lorentzian peaks in the spectrum was accompanied
by the presence of random telegraph switching between a few discrete
levels in the polarization time series.  These features lasted between
a few seconds and at most a few hours showing the transient but
sometimes long-lived nature of the heterogeneity.

Confocal microscopy allows one reconstruct the trajectory of each
particle in a three dimensional colloidal suspension made of several
thousand particles.~\cite{Kegel,Weitz} Using this technique Kegel and
van Blaaderen~\cite{Kegel} found a non-Gaussian distribution of
particle displacements in a dense system of colloidal hard spheres and
they associated this large distribution to the presence of dynamical
heterogeneities.

A visual way of characterizing heterogeneities in the dynamics of a
particle system is to classify the particles according to their
mobility during a chosen time-interval and then study the geometric
and dynamic properties of these subensembles. Even if there is no
unique definition of the mobility of a particle, several reasonable
definitions have been used\cite{Donati99}\footnote{Donati {\it et al}
chose the time $\Delta t$ over which the displacement $|\vec
r_i(t+\Delta t)-\vec r_i(t)|$ is calculated to be the one that yields
the maximum non-Gaussianity of the -- spatial -- van-Hove
correlator~\cite{Donati99}.  This time lies between the end of the
$\beta$-relaxation and the beginning of the $\alpha$-relaxation
The mobility of a particle is then defined as its
maximum displacement during $\Delta t$.}
numerically and experimentally and particular attention has been payed
to the behavior of the two extreme cases of, say, the $5\%$ most
mobile and $5\%$ most immobile particles.

The molecular dynamic simulations of Donati~{\it el
al}~\cite{Donati99} show that at any given instant most of the
particles in a super-cooled liquid oscillate within their cages apart
from $5\%-6\%$ that move quickly along stringy paths on which
particles follow one another.
The mean cluster size of mobile particles\footnote{Clusters of
particles with a chosen mobility are usually defined as the ensemble
of such particles which are closer than the first minimum of the pair
static correlation function.}  increases when lowering the temperature
suggesting the existence of a percolation transition at the estimated
mode-coupling temperature $T_c$ with a divergence of the cluster size
in the thermodynamic limit. The fractal dimension of these clusters is
close to $1.75$. In contrast, the immobile particles, i.e. those with
a small self-displacement, group in rather compact clusters with a
mean cluster size that is quite independent of $T$. The mobility of a
particle is directly related to its potential energy, with the most
mobile (immobile) particles having the highest (lowest) potential
energy.
There is also a correlation between mobility and local composition,
with immobile particles being related to small composition
fluctuations in the mixture.  The time-dependent correlation
functions, restricted to the mobile particles only, have a very
similar structure to the bulk ones with a two-step decay and a cage
effect, only that the life-time of the cage is shorter than on average
and the second decay can be seen in time differences on which the bulk
correlation only reaches the plateau.  Very few particles retain their
mobility after the chosen time-interval, {\it i.e.} the fast or slow
character of a particle is renewed.  The relaxation does not
correspond to ``independent'' volumes relaxing exponentially each with
its own relaxation time.

The organization in mobile and immobile particles has also been
observed in the colloidal dense liquids studied with confocal
microscopy and Weeks {\it el al} showed that, indeed, the typical
cluster size grows when approaching the glass
transition.\cite{Weitz,Weeks1}

Fewer studies of spatial heterogeneities in the glassy phase
exist~\cite{Ogli,Vollmayr,Caetal02,Caetal02-2}.  Vollmayr-Lee {\it et
al} analyzed the structure of the binary Lennard-Jones mixture below
its glass transition~\cite{Vollmayr} using a similar approach to the
one developed by Donati {\it el al}~\cite{Donati99} for the
super-cooled liquid phase. In short, they identified the most mobile
and most immobile particles during a sufficiently short time-interval
and at sufficiently low temperature such that aging effects can be
neglected.
These authors showed that mobile (and immobile) particles tend to be
near each other also in the glassy phase. As expected, mobile
(immobile) particles are placed in regions of lower (higher) density
than average and wider (narrower) cages surround mobile (immobile)
particles. The identity of mobile and immobile particles changes in
time also in this phase.

Experimentally, Courtland and Weeks also found clusters of very mobile
particles in the glassy phase of a colloidal system.~\cite{Weeks2}
(Note that a short coarse-graining in time was implemented in these
measurements in contrast with the previous studies in glycerol where
clustering was not observed~\cite{Miller}.)
Interestingly enough, no evidence for a 
change in the properties of these clusters as the sample
ages has been reported, suggesting that clusters of very mobile 
particles exist even in a very old sample.


The behavior described above can be confronted to the geometric
structure of clusters of spins with similar values of the
coarse-grained two-time correlations in the aging dynamics of a
spin-glass where quenched disorder is very important in determining
the spatial properties of each sample~\cite{Caetal02-2}.  We found
that clusters of negatively valued coarse-grained two-time
correlations that correspond to very fast, under constrained spins,
are well localized. In contrast, clusters of positively valued, slower
spins, are practically space-filling.

\subsection{Intermittency}

Large fluctuations in the time series of a global quantity can be
associated to sudden large rearrangements occurring somewhere in the
sample. Thus, the study of fluctuations in the instantaneous
measurement of a global quantity can be very useful to characterize
the spatial dynamics.

The nature of temporal fluctuations are usually studied via the
frequency dependence of the noise
spectra~\cite{Weiss,noise-SG-Paris,noise-SG,dAnna} (search for $1/f$
noise). This quantity, being the Fourier transform of a temporal
correlation function, only gives information about the second moment
of the fluctuating quantity. Richer information can be extracted from
the full probability distribution. Non-Gaussian distributions of many
interesting observables in glassy systems have been recently reported.
In the rest of this Section we define the measuring procedure and we
recall these experimental results.

In a typical (numerical or real) experiment one monitors a
time-dependent global observable, $O$, and collects the values it
takes with a chosen sampling frequency. A statistical measure involves
many repetitions of the experiment done in identical conditions.
Usually, the reading of the observable on the time grid, $t_k=k \Delta
$, is not strictly local in time but involves a temporal
coarse-graining. In other words, the observable measured on run
$\alpha$ at time $t_k$ is the result of an average over a time window
around $t_k$:
$
\overline O^\alpha(t_k) \equiv n^{-1} \sum_{m=0}^{n-1}
O^\alpha(t_{k-1}+m\delta)
$
with $n\delta =\Delta$. The coarse-graining in time should not
significantly change the main properties of the time-series.  However,
it is clear that the longer the time-interval $\Delta$, keeping
$\delta$ fixed, the smoother the signal. Using an excessive averaging might
erase the effect of rare events and hide the intermittent character of
the relaxation.

Given the time-series for $\overline O^\alpha(t_k)$ one can then check
how the comparison of the signal taken at two-times separated by a
constant lag $\tau$ evolves in time. That is to say, one takes
$C^\alpha(t_k+\tau,t_k) \equiv \overline O^\alpha(t_k) \overline O^\alpha(t_k+\tau) 
$
and follows its evolution as $t_k$ spans the interval
$[t_1,t_2]$.\footnote{Note that this quantity is not a correlation
function since there is no averaging in its definition.}  (In
Fig.~\ref{fig:corr} this procedure corresponds to moving vertically at
a constant value $\tau$ of the temporal axis.)  Choosing $t_1$ very
long one ensures that the averaged bulk correlations, and possibly the
result of each run, decay very slowly in two steps. If $\tau$ does not
go very far beyond $t_1$, for all $t_k$ one explores times and
time-differences that fall on the $\beta$-relaxation. With the
collection of data points $C^\alpha(t_k+\tau,t_k)$, for $\alpha$
ranging from $1$ to the number of runs, and
$k=1,\dots,(t_2-t_1)/\delta$ one can construct a histogram or a
probability distribution. It turns out that its form and its
dependence on the times $t_1$ and $\tau$, yield very interesting
information about the dynamics of the samples.

The ``time-resolved light-scattering technique''~\cite{Luca2} has been
developed with the aim of testing these fluctuations.  Using a
multi-speckle collector, $C^\alpha(t_k+\tau,t_k)$ is calculated as an
average over speckles of the intensity-intensity two-time correlations
in light scattering measurements.  Temporal fluctuations in colloidal
suspensions have been studied in this way. Interestingly enough, this
group has shown that aging samples, such as a concentrated
colloidal gel, have an intermittent dynamics leading to negatively
skewed distributions of the two-time intensity-intensity correlations.
This non-trivial temporal behaviour, once averaged over a short 
time-window, lead to a decay of the
intensity correlation function, $g_2-1$, as the one shown in the right
panel of Fig.~\ref{fig:corr}.


Evidence for intermittency in the {\it global} voltage signal noise in
laponite, a solid-like colloidal glass, and Makrofol DE 1-1 C, a
polymer glass, has been presented recently~\cite{Sergio2}. While the
samples are out of equilibrium and aging, as proven by independent
light scattering measurements in the case of laponite~\cite{Bonn} or
the non-stationarity of the noise spectrum in the case of the polymer
glass, the voltage time series has bursts with very large
amplitude. With sample age these bursts become rarer in the sense that
their amplitude decreases and the time elapsed between two consecutive
ones increases. Eventually, when aging stops the bursts disappear from
the noise signal. In order to better characterize the signal Buisson
{\it et al}~\cite{Sergio2} constructed a waiting-time dependent {\sc
pdf}, $\rho(V)$, by dividing time in intervals of duration $\tau$.
The {\sc pdf}s are clearly negatively skewed (intermittent noise) at
short times and they progressively approach a Gaussian distribution
when the system approaches equilibrium.

Using confocal microscopy
Courtland and Weeks demonstrated that the averaged (over all
particles) displacement between a waiting-time and a subsequent time
is also intermittent leading to a non-Gaussian
distribution.~\cite{Weeks2} It is interesting to note that these
experiments show that significant fluctuations occur even if the
time-lag $\tau$ is very short compared to $t_w$.


\section{Heterogeneous aging dynamics: theory} 
\label{theory}

Recently, we constructed a framework for the study of fluctuations in
the nonequilibrium relaxation of glassy systems with and without
quenched disorder.~\cite{Chetal02,Caetal02,Chetal04,Caetal02-2}
In these articles we studied coarse-grained local correlators obtained
for a given noise or experimental realization, and the fluctuations of
two-time global quantities in finite-size systems, but we did not
study the temporal fluctuations, leading to intermittent or continuous
dynamics as found experimentally~\cite{Sergio2,Luca2}.  Here we
present preliminary results concerning the behavior of the latter
fluctuations in spin-glass models~\cite{Chetal04}.  Focusing on the
fluctuations of the coarse-grained local two-time quantities we
predicted constraints on their
distributions.~\cite{Chetal02,Caetal02,Chetal04,Caetal02-2} In particular we
showed that locally defined correlations and responses are connected
by a generalized local out-of-equilibrium fluctuation-dissipation
relation. We argued that large-size heterogeneities in the age of the
system should survive in the long-time limit. A symmetry of the
underlying theory, namely invariance under reparametrizations of the
time coordinates, is at the basis of these results.  We established a
connection between the probabilities of spatial distributions of local
coarse-grained quantities and the theory of dynamic random manifolds.
We defined and discussed the behavior of a two-time dependent
correlation length from the spatial decay of the fluctuations in the
two-time local functions.  For concreteness, we presented numerical
tests performed on disordered spin models in finite and infinite
dimensions.  We characterize the fluctuations in the system in terms
of clusters of coarse-grained sites with similar properties, similarly
to what has been done when studying super-cooled liquids and glasses
numerically~\cite{Donati99} and
experimentally~\cite{Weitz,Weeks1,Weeks2}.  Finally, we explained how
these ideas can be applied to the analysis of the dynamics of other
glassy systems that can be either spin models without disorder or
atomic and molecular glassy systems.  In this Section we review part
of these results and we present some more recent studies of the form
of the probability distribution functions of the two-time correlation
functions.

\subsection{Distinction between fluctuations: spatial, dynamical
and finite size}

Let us define several two-time auto-correlations that differ in how
they have been averaged.  For concreteness, we consider a cubic system
with fixed volume $V=L^d$.

\subsubsection{Temporal fluctuations in global measurements}
\label{temporal-def}

\begin{figure}
\begin{center}
\begin{tabular}{c}
\includegraphics[height=4.3cm]{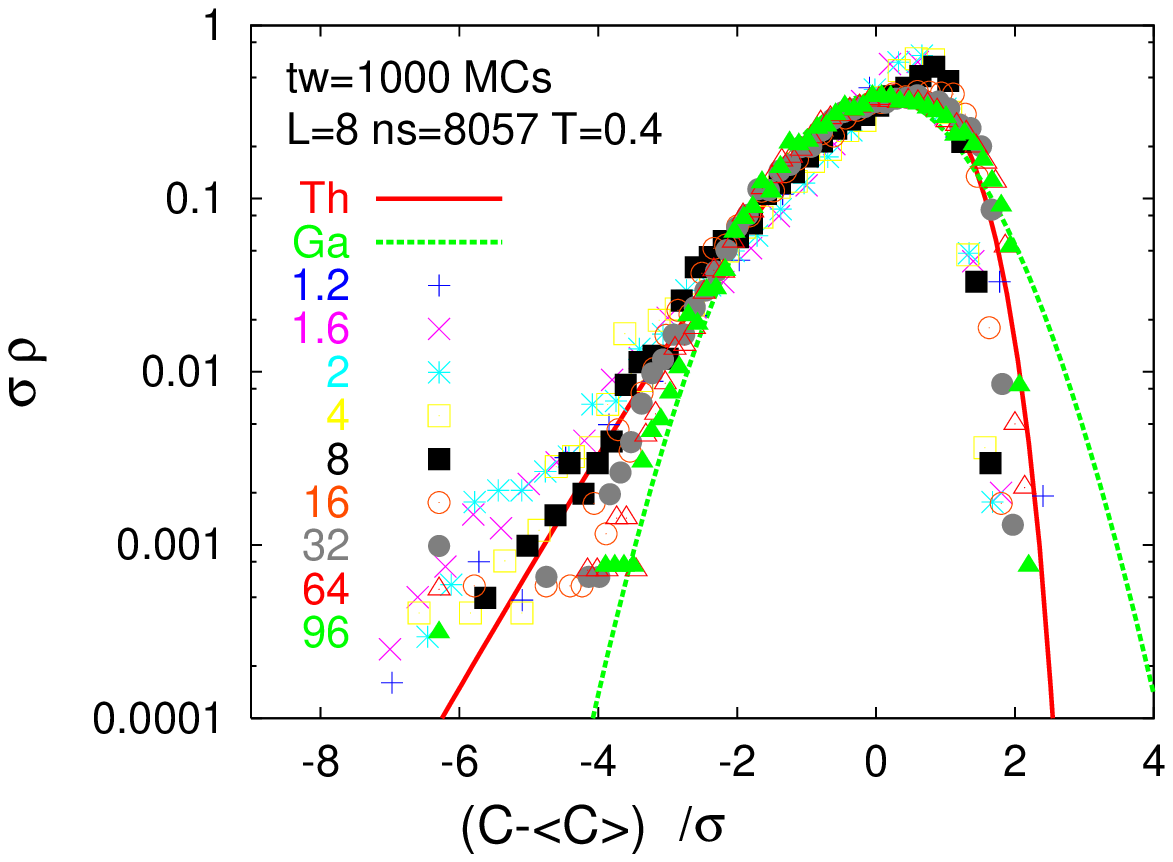}
\includegraphics[height=4.3cm]{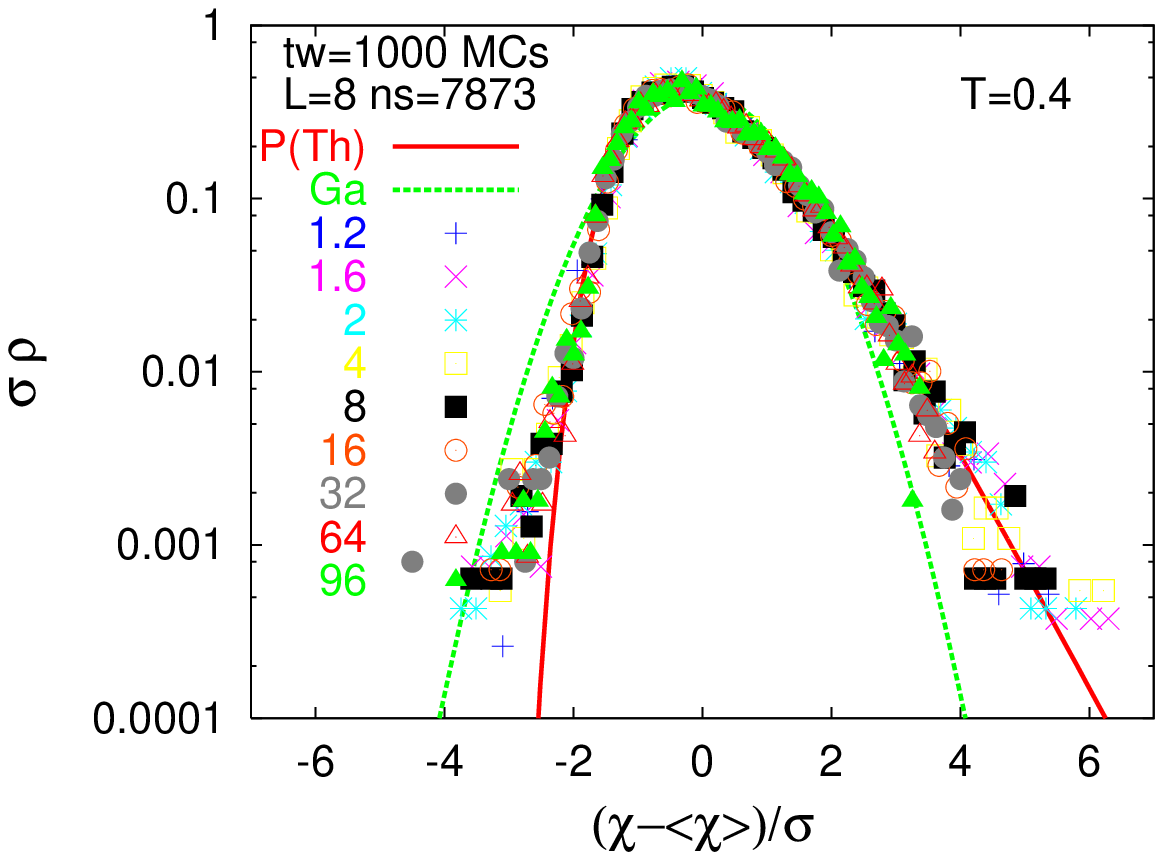}
\end{tabular}
\end{center}
\caption{
\label{fig:Gumbel}
Scaled probability distribution of the mesoscopic fluctuations in the
global correlation, $N^{-1} \sum_{i=1}^N s_i(t) s_i(t_w)$ (left), and
susceptibility, $N^{-1} \sum_{i=1}^N \delta
s_i(t)/\epsilon_i(t_w)|_{\epsilon=0}$ (right), of the $3d${\sc ea}
model with linear size $L=8$ at $T=0.4<T_c$ obtained form Montecarlo
simulations.~\cite{Chetal04} $\sigma$ and $\langle C\rangle$ (and
$\langle \chi\rangle$) are the variance and average of the raw {\sc
pdf}, $\rho$, at each pair of times.  A temporal coarse-graining of
the spin configuration has been implemented, with a coarse-graining
time equal to $t/10$.  Different sets of points correspond to
different values of $t/t_w$ that are given in the key. The
waiting-time is $t_w=10^4$ {\sc mc}s. $n_s$ is the number of runs,
each one leading to a point in the generation of the histogram. The
red line is a generalized Gumbel {\sc pdf}\cite{BHP} with $b=\pi/2$,
see Eq.~(\ref{gumbel}). The green line is a Gaussian {\sc pdf} with the
same variance. See Ref.~\cite{Chetal04} for more details on the 
functional form of these {\sc pdf}s.} 
\end{figure}

In numerical simulations of Ising spin models a useful choice for the
observable is $O^\alpha=N^{-1/2} \sum_{i=1}^N s^\alpha_i$.  Note that
$C^\alpha(t_k+\tau, t_k)=N^{-1}\sum_i s^\alpha_i(t_k+\tau)
s^\alpha_i(t_k)$ (since crossed terms cancel) and it is normalized to
one at equal times.  Interestingly enough, the {\sc pdf} of $C^\alpha$
takes a very similar form~\cite{Chetal04} to the one found in the light
scattering measurements of some colloidal suspensions~\cite{Luca2}
(see Fig.~\ref{fig:Gumbel}).  For infinitesimal short lag-times $\tau$
one expects a Gaussian distribution of fluctuations but this regime is
not of easy access neither numerically nor experimentally.  For longer
but still short lags that correspond to the end of the
$\beta$-relaxation one finds a negatively skewed distribution with a
rather long tail towards small values of $C$.  For much longer lags
such that one enters deeply in the structural relaxation regime the
distribution gets more symmetric and eventually becomes Gaussian
again~\cite{Luca2,Chetal04}.

The (two-time dependent) negatively skewed {\sc pdf} can be rather
well described with a {\it generalized Gumbel
distribution}:~\cite{Luca2,Chetal04}
\begin{equation}
\rho_b(x) = \frac{|b|}{\Gamma(b)} e^{b\ln b} e^{b
\left(\alpha(x-x_o)-e^{\alpha(x-x_o)}\right)}
\; .
\label{gumbel}
\end{equation}
When $b=1$ this is the usual Gumbel distribution of the first
kind. When $b=\pi/2$ Bramwell, Holdsworth and Pinton~\cite{BHP} found
that it describes very accurately the fluctuations in the energy
injected in a close turbulent flow at fixed Reynolds number, it
characterizes exactly the fluctuations in the $2d$ {\sc xy} model
close to its critical line and it also describes many other types of
fluctuations in very different critical systems.  R\'acz {\it et al}
discussed these distributions in connection with the roughness of
random surfaces and $1/f$ noise~\cite{Zoltan,Zoltan2}.
In Sect.~\ref{manifold} we shall discuss a justification to use this
distribution based on an effective random manifold model that we
proposed controls the fluctuations in the temporal and spatial
correlations~\cite{Chetal02,Caetal02,Chetal04,Caetal02-2}. We shall argue that
it might be possible to scale the {\sc pdf}'s for all $t_w$ and $t$
using modified Gumbel distributions as in (\ref{gumbel}) with a
parameter $b$ that is real and smoothly two-time dependent for aging
problems.

\subsubsection{Spatial fluctuations}

In Sect.~\ref{exps} we described the observation of local dynamic
heterogeneities in some glassy systems. These studies focused on
single particle measurements. In order to derive a theoretical model
including these effects it is more convenient to introduce a spatially
coarse-grained description. With this aim let us divide the system
with volume $V=L^d$ into $n$ subsystems with volume $v=\ell^d$. One
can then compare the configuration of the system within the same local
volume at two different times. This defines what we call a two-time
coarse-grained local correlation (even if, strictly, this is not a
correlation). Again, this is a single run quantity. To obtain better
probability distributions one can repeat the measurement and get
$L/\ell$ data points for each run.

For an Ising spin model on a lattice we have
$C^{cg}_i(t,t_w) \equiv {v}^{-1} \sum_{j\in v_i} s_j(t) s_j(t_w)$,
where $v_i$ is the coarse-graining volume surrounding the site $i$
that equals the number of spins within this volume. For simplicity,
one takes it to be the same for all sites in the sample. We omitted
the index $\alpha$ labelling runs to make the notation more
compact. The local coarse-grained correlation is normalized to one at
equal times.  Another interesting example is the case of a particle
system in a $d$ dimensional continuum space. In this case, the density
fluctuations are probably the simplest and most interesting
observables. One then has
$C^{cg}_i(t,t_w) \equiv (\rho_i(t)-\langle \rho(t) \rangle) 
(\rho_i(t_w)-\langle \rho(t_w) \rangle) $
that one can normalize to be equal to one at equal
times~\cite{KA-aging}.

If we use a sufficiently large coarse-graining volume $v$, the local
correlations should maintain the more salient qualitative features of
the decay of their bulk partners.~\cite{Caetal02,Caetal02-2} Thus, the
local correlations should decay quickly towards a {\it local}
Edwards-Anderson parameter, $
q^i_{\sc ea} \equiv \lim_{t-t_w\to\infty} \lim_{t_w\to\infty}C_i(t,t_w) 
$, and then slowly below this value. If the structure of the global
correlation is preserved at the local level, the first step of the
relaxation should be stationary whilst the second one could be
waiting-time dependent.  Even if site-to-site fluctuations in
$q^i_{\sc ea}$ are in principle possible,
we expect them to be erased by using a 
sufficiently large coarse-graining volume $v$, 
and have $q_{\sc ea}^i=q_{\sc ea}$. 
The decay below the 
site-independent plateau should then carry all the information 
about the spatial fluctuations. 
Extending the scaling of monotonically decaying 
two-time correlations within a correlation-scale,~\cite{Cuku94}
and using the fact that we expect that 
the effective dynamic action describing
the dynamics for well-separated times in the structural
relaxation regime  becomes time-reparametrization 
invariant,\cite{Chetal02}
we argued that the local coarse-grained correlations in the slow
regime should scale as~\cite{Caetal02-2}
\begin{equation}
C_{slow}^i(t,t_w) = f_{slow}\left(\frac{h_{slow}^i(t)}{h_{slow}^i(t_w)}\right) =
f_{slow}\left(e^{\phi_{slow}^i(t)-\phi_{slow}^i(t_w)}\right)
\; .
\label{eq:localh}
\end{equation}
for all sites in the sample. In what follows we omit the subindex
$slow$ since we always refer to this regime.  This proposal implies
that different sites can evolve on totally different time-scales or
there can be small local fluctuations in the scaling function $h_i(t)$
with respect to the global scaling function $h(t)$, see
Eq.~(\ref{global-C}), such that $h_i(t) = h(t)+\delta h_i(t)$, with
$|\delta h_i(t)| \ll h(t)$.
We expect the latter behavior to be more common since rapid variations
in $C_i$ -- in space and time -- should be unfavored by some stiffness
making these variations costly.  Note that the value of the
correlations on two sites $i,j$ may cross each other as a function of
time though we expect this to happen at very short time-differences only.

\subsubsection{Finite size fluctuations}

Another interesting kind of fluctuations are those associated to the
finite (say mesoscopic) size of a sample. The idea is, simply, to
study the fluctuations in the reading of a global quantity between
different repetitions of the same experiment.

\subsection{Sigma model}
\label{sigma}

We recently studied the symmetry properties of the dynamic action for
the aging dynamics of $d$-dimensional spin-glasses~\cite{Chetal02}. We
derived a disorder-averaged dynamic generating functional that is a
path integral over local coarse-grained two-time functions. The aim of
this work was to show that for longer and longer waiting-times the
action for the slow parts of the two-point functions that describe the
structural relaxation progressively acquires a global symmetry (a zero
mode develops) that is the invariance under identical transformations
of time $t\to h(t)$ for all local functions. (In this proof we assumed
that there is a {\em local} separation of time scales and that the
system is causal.) Note that this means that the effective slow system
becomes critical and the arguments discussed in the Introduction
apply. The contributions to this action having an origin in the fast
part of the relaxation (the approach to the plateau) act as a weak
symmetry breaking field that selects a scaling function $h(t)$ that
determines the dynamics of the global two-time functions. Now,
fluctuations can be of two types. Either the external function $f$
(longitudinal) or the internal function $h$ (transverse) fluctuate
from site to site. The symmetry of the action indicates that the
former are less favorable while the latter cost the lowest action.
Thus, the more favorable fluctuations correspond to spatially-varying
time reparametrizations, $h(\vec r,t)=h(t)+\delta h(\vec r,t) $.  We
expect that a stiffness will be generated once the two-time quantities
are coarse-grained, making sharp variations in $\delta h(\vec r,t)$
difficult to achieve.  One of the main consequences of this analysis
is that the coarse-grained two-time correlations and susceptibilities
should behave as described in Sect.~\ref{Teff}.

\subsection{Effective random surface theory}
\label{manifold}

Unfortunately, we are not able to study the exact dynamic generating
function for a glassy problem in finite dimensions.  Still, guided by
symmetry constraints we can propose a phenomenological effective
action for the slow modes and derive generic conclusions from
it~\cite{Chetal02,Caetal02,Chetal04,Caetal02-2}.  Indeed, we expect the 
actual
effective action for the slow modes to become invariant invariant
under global time reparametrizations in the long waiting-time
limit. Therefore, any phenomenological effective action we can propose
must have this symmetry.  We then select from all possible
time-reparametrization invariant terms the ones that we estimate are
the leading one in the long times limit. These include a term
penalizing rapid variations (in space and time) of the coarse-grained
local two-time functions that involves a gradient and time-derivatives
and other ``potential'' terms that are dictated by the interactions in
the system. The saddle-point is time-dependent but uniform in space so
it is completely determined by the second type of terms.  It fixes,
for example, the function $f$ and the global scaling function $h$ in
Eq.~(\ref{global-C}).  The gradient term is very important to
determine the fluctuations. Defining a proper time ${\cal T}=\ln
h(t)$, and parametrising the fluctuations with the fields $\phi(\vec
r,{\cal T})$, we proposed~\cite{Caetal02,Caetal02-2,Chetal04} that the action
controlling the fluctuations in generic systems with slow dynamics is
a random surface theory in $d$ dimensions. This proposal allowed us to
derive the time-scalings of the {\sc pdf}'s of local
correlations~\cite{Caetal02} and we recently exploited 
it~\cite{Chetal04} to justify the Gumbel-like form of the {\sc pdf}'s,
see Fig.~\ref{fig:Gumbel}.

\subsection{Two-time dependent correlation length}

The fluctuating quantities in the theory are two-time ``fields''
associated to the coarse-grained local correlations and responses.  In
the asymptotic limit, the time-reparametrization invariance implies a
true Goldstone, or zero mass, mode. Therefore, the spatial
correlations in the fluctuations should show a power law decay $\sim
1/r$ in $3d$. However, for any finite time, this symmetry is
explicitly broken by irrelevant terms that play the role of symmetry
breaking fields and the Goldstone mode has a small mass. In a
simulation we expect to find a finite correlation length for the the
coarse-grained local two-time quantities. A convenient measure of the
spatial correlations in the fluctuations is
\begin{equation}
B(r;t,t_w) \equiv \frac{\tilde A(r;t,t_w)-\tilde A_\infty(t,t_w)}
{\tilde A(r=0;t,t_w)-\tilde A_\infty(t,t_w)}
\label{EQ:B_def}
\;\;\;\;\;\;\;\;\;\;
\mbox{with}
\;\;\;\;\;\;\;\;\;\;
\tilde A(r;t,t_w)
\equiv
\left[\frac{1}{N}\sum_i C_{{\vec r}_i}(t,t_w)\; 
C_{{\vec r}_i+\vec r}(t,t_w) \right] 
\end{equation}
Fixing the relation between $t$ and $t_w$ to have, say, a given global
correlation, one should find a correlation length $\xi(t,t_w)$ that
increases monotonically for increasing $t_w$.  We have recently
studied this quantity for a finite $d$ spin-glass
model~\cite{Caetal02-2} and found that $\xi(t,t_w)$ takes very small
values for the times reachable numerically showing that the system is
very far from its true asymptotic regime. We refer the reader to this
reference for more details on this quantity.

\subsection{Fluctuations in the local effective temperatures}
\label{Teff}

The analytic solution to mean-field disordered spin 
models~\cite{Cuku93,Cuku94,Cu02} includes 
rather simple modifications of the equilibrium {\sc fdt} relations
between induced and spontaneous fluctuations.
This new relation reads 
\begin{equation}
\lim_{t_w\to\infty \; C(t,t_w)={\tt C}}\chi(t,t_w) = \tilde \chi({\tt C}) 
\; ,
\end{equation}
where $\chi(t,t_w)$ is a linear response function that has been
integrated over time from $t_w$ to $t$ and $C$ is its conjugated
correlation. $\tilde \chi$ is a model-dependent function that relates
this two-time quantities. One finds that for glassy systems, $\tilde
\chi(C)=1/T(1-C)$, with $T$ the temperature of the environment, for
values of the correlations that are above the plateau, and $\tilde
\chi(C)$ takes a different linear form for values of the correlation
below the plateau. For theoretical spin-glass models the second part
of the function is curved.
This phenomenon has been tested numerically~\cite{Cu02} and 
experimentally~\cite{Didier,Sergio1,Sergio2,Tomas} in 
a very large variety of systems with slow dynamics.


Let us now define a generalized {\it local} fluctuation-dissipation
relation ({\sc fdr}) via the limit~\cite{Caetal02-2}
\begin{equation}
\lim_{t_w\to\infty\; C_i(t,t_w)={\tt C}_i}
\chi_i(t,t_w) =  \tilde\chi_i({\tt C}_i)
\; 
\label{localchiiCi}
\end{equation}
that exists 
if the local correlations are monotonic functions of $t$ for fixed
$t_w$.
Based on thermometric arguments~\cite{Cukupe97}, we associate the
variation of $\tilde\chi_i$ with respect to $C_i$ to a local
effective-temperature.  The question now arises as to whether the
fluctuations in $C_i$ and $\chi_i$, and the local
effective-temperatures, are independent or whether they satisfy
certain relations.

\begin{figure}
\begin{center}
\begin{tabular}{c}
\hspace{-0.25cm}
\vspace{-0.5cm}
\includegraphics[height=3.8cm]{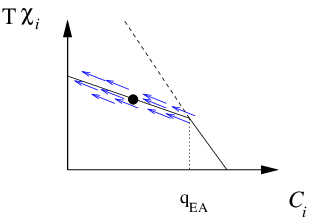}
\hspace{-1cm}
\includegraphics[height=3.8cm]{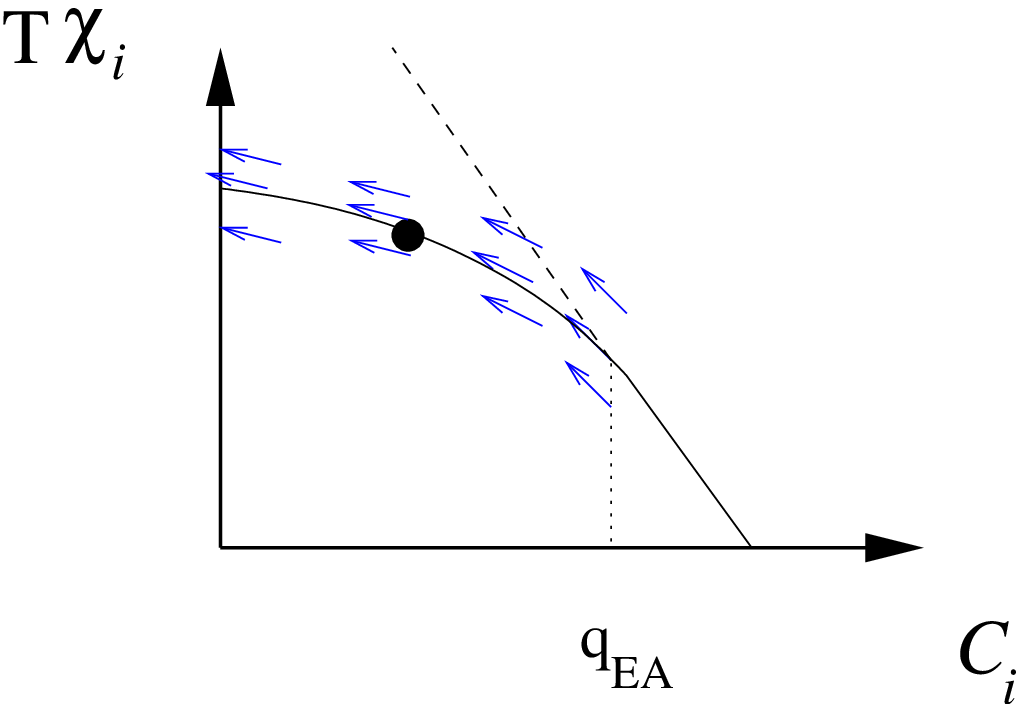}
\hspace{-1cm}
\includegraphics[height=5.7cm]{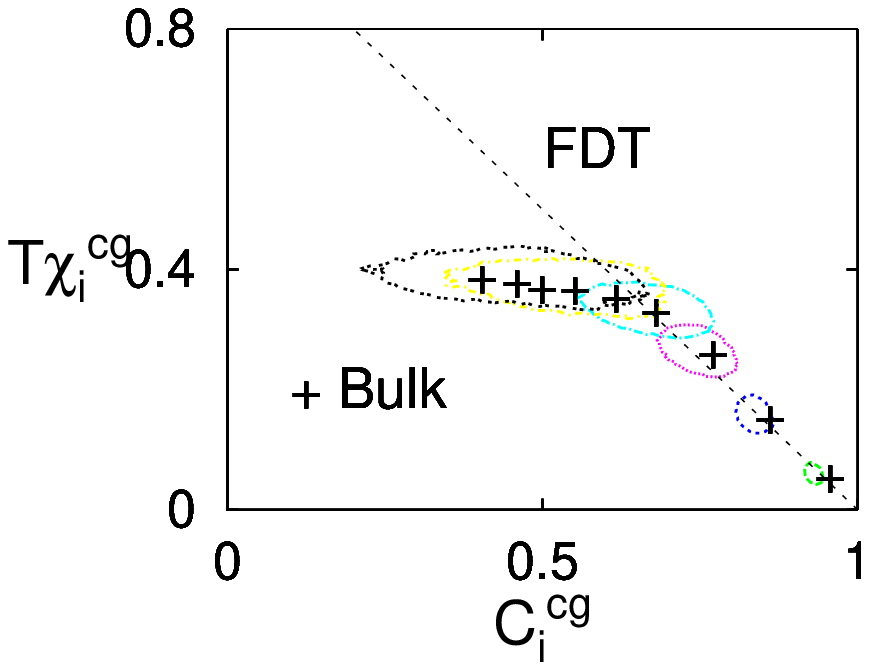}
\end{tabular}
\end{center}
\caption{
\label{fig:chiCnos1step}
Left: Sketch of the fluctuations in a system with a single correlation
scale below $q_{\sc ea}$. The symmetry argument implies that
all pairs are concentrated along the global straight line with little
dispersion perpendicular to it. The arrows have the direction of the
second slope in the global $\tilde \chi(C)$ curve.  Center: A system
with a sequence of correlation scales below the global $q_{\sc ea}$.
All point positions are constrained to be near the global
$\tilde\chi(C)$ curve for $C < q_{\sc ea}$, and their velocities are
forced to be parallel to the global $\tilde\chi(C)$ curve for $C <
q_{\sc ea}$.  The velocities fluctuate from site to site, but since
they are determined by the global curve, they are identical for all
sites with the same value of the local correlation $C_i$.  In both
cases the dashed line continues the {\sc fdt} line and the black dot
indicates the location of the values for the susceptibility and
correlation averaged over the distribution.  Right: results from a
Montecarlo simulation of the $3d$ {\sc ea} model below its critical
temperature~\cite{Caetal02,Caetal02-2}, the coarse-graining volume is
$v=\ell^3=13^3$ and $L=32$. The crosses correspond to the averaged
values and the lines are the projection of one selected contour level
for the same pair of times.  }
\end{figure}

In the $C-T\chi$ plane, the soft (massive) modes discussed in
Sect.~\ref{sigma} correspond to displacements along (transverse to)
the global $\chi(C)$ curve. Thus, the spatial fluctuations in the {\sc
fdr} should be such that the projection of the joint {\sc pdf},
$\rho(C_i,\chi_i)$, computed at two fixed times $t$ and $t_w$
concentrates along the global $\chi(C)$ curve.  In
Fig.~\ref{fig:chiCnos1step} we show a sketch of this quantity.  Given
a pair of times $t_w\leq t$, we depict the $N$ points $(C_i(t,t_w),T
\chi_i(t,t_w))$ with arrows that represent the velocity of the points
[{\it i.e}, the rate at which the $(C_i,T \chi_i)$ positions change as
one changes $t$], and are located at their position in the $C-T\chi$
plane.
In the same plots we draw the parametric
plot for the global $T \tilde \chi(C)$ constructed as
usual~\cite{Cuku94}: for a fixed $t_w$ we follow the evolution of the
pairs $(C,T\chi)$ as time $t$ evolves from $t=t_w$ to $t\to \infty$.
We scale the y-axis by temperature to work with dimensionless
variables.

The scaling in Eq.~(\ref{eq:localh}) means that different sites might
evolve on different time-scales and hence have their own effective
temperature.
Based on the analytic study of the fluctuations in a disordered spin
model~\cite{Chetal02} we expect 
the local responses and correlations to be constrained to follow the 
global curve, {\it i.e.} 
\begin{equation}
\tilde\chi_i(C_i) = \tilde\chi(C_i) 
\; ,
\label{globalconstrain}
\;\;\;\;\;\;\;\;\;\;\;\;
\mbox{and}
\;\;\;\;\;\;\;\;\;\;\;\;
-\beta_i^{\sc eff}(C_i) \equiv \frac{d\tilde\chi_i(C_i)}{dC_i} =
\frac{d\tilde\chi(C_i)}{dC_i} =
\left. \frac{d\tilde\chi(C)}{dC}\right|_{C=C_i}
\; .
\label{localTidep}
\end{equation}
If there are only two time-scales for the decay of the global 
correlation, below $q_{\sc ea}$
$\tilde\chi(C)$ is linearly dependent on $C$ and this equation yields 
$
\beta_i^{\sc eff} = \beta^{\sc eff}
$,
for all sites, and values of $C_i<q_{\sc ea}$.  If, on the contrary,
the global correlation decays in a sequence of scales and
$\tilde\chi(C)$ is not a linear function of $C$, one has fluctuations
in the local effective temperature due to the fluctuations in $C_i$.
This behavior is sketched in Fig.~\ref{fig:chiCnos1step}.  for a
system with two global correlation-scales (left) and a system with a
sequence of global correlation-scales (right).

If the two times $t$ and $t_w$ are close to each other, in such a way
that the global correlation between them lies above $q_{\sc ea}$, the
global correlator and the global linear integrated response are
stationary and related by the {\sc fdt}. In this regime of times we
also expect the local quantities to be linked by the {\sc fdt}. Note
that the arguments based on time reparametrization invariance do not
apply to these short time-differences since the relaxation is not slow
here.  We have verified numerically~\cite{Caetal02} that the
fluctuations of the local coarse-grained correlations and responses
are concentrated rather spherically around the global value
$\chi=(1-C)/T$ when $C>q_{\sc ea}$.

\section{Conclusions}

Experimental, numerical and theoretical effort is presently devoted to
elucidate the role played by fluctuations in the dynamics of
super-cooled liquids and glasses.  From our point of
view~\cite{Chetal02,Caetal02,Chetal04,Caetal02-2} this problem can be
tackled theoretically by using effective sigma-models constrained by
the symmetries in the problem and, most importantly, by the expected
asymptotic time-reparametrization invariance.  This approach allowed
us to make precise predictions on the form of the individual and joint
probability distribution functions that can be put to numerical and
experimental tests. The results of such measurements will yield very
useful feedback to fix the terms in these effective models.
Presumably, some kind of ``universality'' will apply and one can
expect to find that a sigma-model represents the behaviour of many
different classes of glasses.

\acknowledgments 
 
LFC is research associate  at ICTP Trieste Italy.
I wish to specially thank H. E. Castillo, C. Chamon, P. Charbonneau,
J. L. Iguain, M. P. Kennett, D. Reichman and M. Sellitto for our
collaboration on the recent results that I discussed in this
article. I also want to thank L. Cipelletti, D. H\'erisson and
M. OcioI for very interesting discussions on their experimental
work as well as L. Berthier and J. P. Garrahan. 
I acknowledge financial support from the ACI project
"Optimisation algorithms and quantum disordered systems", the
Guggenheim Foundation, the National Science Foundation under Grant
No. PHYS99-07949, and I thank the KITP at the University of
California at Santa Barbara, USA and the ICTP, Trieste, Italy for
hospitality during the preparation of this article.


\end{document}